\documentclass[final,authoryear,3p,times]{elsarticle}

\usepackage{amssymb}
\usepackage{epstopdf}
\usepackage{lmodern}
\usepackage[latin9]{inputenc}
\usepackage[T1]{fontenc}
\usepackage{color}
\usepackage{float}
\usepackage{tabularx}
\usepackage{breqn}
\usepackage{pgfplots}
\pgfplotsset{compat=1.5}
\newcommand{\mbf}{\mathbf}

\journal{International Journal of Solids and Structures}

\newcommand{\hsp}[1]{\textcolor{blue}{[\textit{hsp}: #1]}}

\begin{document}

\begin{frontmatter}



\title{Computational Modeling of Electro-Elasto-Capillary Phenomena in Dielectric Elastomers}


\author[ss]{Saman Seifi}
\ead[ss]{samansei@bu.edu}
\address[ss]{Department of Mechanical Engineering, Boston University, Boston, MA 02215}
\author[hsp]{Harold S. Park}
\ead[hsp]{parkhs@bu.edu}
\address[hsp]{Department of Mechanical Engineering, Boston University, Boston, MA 02215}

\begin{abstract}

We present a new finite deformation, dynamic finite element model that incorporates surface tension to capture elastocapillary effects on the electromechanical deformation of dielectric elastomers.  We demonstrate the significant effect that surface tension can have on the deformation of dielectric elastomers through three numerical examples:  (1) surface tension effects on the deformation of single finite elements with homogeneous and inhomogeneous boundary conditions; (2) surface tension effects on instabilities in constrained dielectric elastomer films, and (3) surface tension effects on bursting drops in solid dielectrics.  Generally, we find that surface tension creates a barrier to instability nucleation.  Specifically, we find in agreement with recent experimental studies of constrained dielectric elastomer films a transition in the surface instability mechanism depending on the elastocapillary length.  The present results indicate that the proposed methodology may be beneficial in studying the electromechanical deformation and instabilities for dielectric elastomers in the presence of surface tension. 

\end{abstract}


\end{frontmatter}


\section{Introduction} 



Dielectric elastomers (DEs) have attracted significant attention in recent years as a soft and flexible actuation material~\citep{carpiSCIENCE2010,brochuMRC2010,biddissMEP2008}.  The salient characteristic of DEs is that if sandwiched between two compliant electrodes that apply voltage across its thickness, the DE can exhibit both significant thinning and in-plane expansion, where the in-plane expansion can often exceed several hundred percent~\citep{keplingerSM2012}.  The ability to undergo such large deformations has led to DEs being studied for both actuation-based applications, including artificial muscles and flexible electronics, and also for generation-based applications and energy harvesting~\citep{carpiSCIENCE2010,brochuMRC2010,mirMT2007}.

Detailed studies on the mechanics of DEs began about 15 years ago with the seminal experimental work of~\citet{pelrineSAA1998,pelrineSCIENCE2000}.  Since then, there have been many experimental~\citep{foxJMPS2008,keplingerPNAS2010,kofodJIMSS2003,kofodSAA2005,peiSPIE2004,planteIJSS2006,planteSAA2007,planteSMS2007,schlaakSPIE2005,wisslerSAA2007a,zhangSPIE2004,chibaSPIE2008,wangPRL2011a,wangAM2012,wangNC2012}, theoretical~\citep{suoJMPS2008,suoAMSS2010,goulbourneJAM2005,dorfmannAM2005,dorfmannJE2006,mcmeekingJAM2005,patrickSAA2007,planteSAA2007,planteSMS2007,wisslerSMS2005}, and recently a small number of computational studies~\citep{parkIJSS2012,parkSM2013,parkCMAME2013,zhouIJSS2008,zhaoAPL2007,vuIJNME2007,wisslerSMS2005,buschelIJNME2013,khanCM2013,henannJMPS2013,liSMS2012} aimed that identifying the mechanisms that have the largest impact on the nonlinear dynamical behavior and failure mechanisms of DEs.  A summary of recent developments on electromechanical instabilities in DEs has been given by~\citet{zhaoAPR2014}.  

Concurrently, researchers have for many years studied the forces exerted by fluids at rest upon solids, which are known as surface tension, or elastocapillary forces.  While the best-known example of surface tension is likely that of deforming liquid droplets, there has been interest in using it to deform solid structures, see the reviews of~\citet{romanJPCM2010} and~\citet{liuAMS2012}.  Specifically, there has recently been interest in using elastocapillary forces to deform \emph{soft} structures in controllable or unique ways, since for these systems the elastocapillary number, which is defined as $\gamma/\mu l$, where $\gamma$ is the surface tension, $\mu$ is the shear modulus and $l$ is a characteristic length, is close to unity, implying that elastocapillary effects can be substantial for these soft materials.  

While elastocapillary effects have been extensively studied in soft materials, its effect on soft materials like DEs that deform when subject to an electric field, is a relatively unknown phenomenon.  For example,~\citet{wangPRE2013} performed interesting experiments of electrostatically deforming a constrained DE in a liquid solution, and showed that the instability mechanism of the surface could be tuned depending on the value of the surface tension.  Furthermore,~\citet{pineiruaSM2010} showed how elastocapillary origami could be developed by coupling surface tension with electric fields to deform a liquid droplet surrounded by a thin sheet of PDMS.  Overall, this discussion makes clear that there may be potential in using surface tension as an additional degree of freedom to introduce new and interesting deformation mechanisms in electroactive polymers like DEs, and furthermore that the computational tools needed to investigate such phenomena are currently lacking.  

Therefore, the objective of the present work is to present a finite element (FE) model for DEs accounting for the effects of surface tension, such that electro-elasto-capillary phenomena in DEs can be computationally investigated.  We pay particular interest to those instances where surface tension couples to and impacts known electromechanical instabilities that dielectric elastomers are known to undergo, specifically snap-through instability~\citep{pelrineSCIENCE2000}, surface creasing and wrinkling~\citep{wangPRL2011a,wangAM2012,wangPRE2013}, and bursting drops in solid dielectrics~\citep{wangNC2012}.

\section{Background: Nonlinear Electromechanical Field Theory}

The numerical results we present in this work are based upon a FE discretization of the electromechanical field theory proposed by Suo and co-workers~\citep{suoJMPS2008,suoAMSS2010}.   In this field theory at mechanical equilibrium, the nominal stress $S_{iJ}$ satisfies the following (weak) equation:
\begin{equation}\label{eq:suo1} \int_{V}S_{iJ}\frac{\partial\xi_{i}}{\partial X_{J}}dV=\int_{V}\left(B_{i}-\rho\frac{\partial^{2}x_{i}}{\partial t^{2}}\right)\xi_{i}dV+\int_{A}T_{i}\xi_{i}dA,
\end{equation}
where $\xi_{i}$ is an arbitrary vector test function, $B_{i}$ is the body force per unit reference volume $V$, $\rho$ is the mass density of the material and $T_{i}$ is the force per unit area that is applied on the surface $A$ in the reference configuration.  

For the electrostatic problem, the nominal electric displacement $\tilde{D}_{I}$ satisfies the following (weak) equation:
\begin{equation}\label{eq:suo2} -\int_{V}\tilde{D}_{I}\frac{\partial\eta}{\partial X_{I}}dV=\int_{V}q\eta dV+\int_{A}\omega\eta dA,
\end{equation}
where $\eta$ is an arbitrary scalar test function, $q$ is the volumetric charge density and $\omega$ is the surface charge density, both with respect to the reference configuration.  It can be seen that the strong form of the mechanical weak form in (\ref{eq:suo1}) is the momentum equation, while the strong form of the electrostatic weak form in (\ref{eq:suo2}) is Gauss's law.  

As the governing field equations in (\ref{eq:suo1}) and (\ref{eq:suo2}) are decoupled, the electromechanical coupling occurs through the material laws.  The hyperelastic material law we adopt here has been utilized in the literature to study the nonlinear deformations of electrostatically actuated polymers; see the works of~\citet{vuIJNME2007}, and~\citet{zhaoAPL2007}.  Due to the fact that the DE is a rubber-like polymer, phenomenological free energy expressions are typically used to model the deformation of the polymer chains.  In the present work, we will utilize the form~\citep{vuIJNME2007,zhaoAPL2007}
\begin{equation}\label{eq:de1} W(\mbf{C},\tilde{\mbf{E}})=\mu W_{0}-\frac{1}{2}\lambda(\ln{J})^{2}-2\mu W_{0}(3)\ln{J}-\frac{\epsilon}{2}JC_{IJ}^{-1}\tilde{E}_{I}\tilde{E}_{J},
\end{equation}
where $W_{0}$ is the mechanical free energy density in the absence of an electric field, $\epsilon$ is the permittivity, $J=\det(\mbf{F})$, where $\mbf{F}$ is the continuum deformation gradient, $C_{IJ}^{-1}$ are the components of the inverse of the right Cauchy-Green tensor $\mbf{C}$, $\lambda$ is the bulk modulus and $\mu$ is the shear modulus.  

We model the mechanical behavior of the DE using the Arruda-Boyce rubber hyperelastic function~\citep{arrudaJMPS1993}, where the mechanical free energy $W_{0}$ in (\ref{eq:de1}) is approximated by the following truncated series expansion,
\begin{eqnarray}\label{eq:de2} W_{0}(I_{1})=\frac{1}{2}(I_{1}-3)+\frac{1}{20N}(I_{1}^{2}-9)+\frac{11}{1050N^{2}}(I_{1}^{3}-27) \\ \nonumber
+\frac{19}{7000N^{3}}(I_{1}^{4}-81)+\frac{519}{673750N^{4}}(I_{1}^{5}-243),
\end{eqnarray}
where $N$ is a measure of the cross link density, $I_{1}$ is the trace of $\mbf{C}$, and where the Arruda-Boyce model reduces to a Neo-Hookean model if $N\rightarrow\infty$.  We note that previous experimental studies of~\citet{wisslerSAA2007a} have validated the Arruda-Boyce model as being accurate for modeling the large deformation of DEs.  

\section{Finite Element Formulation}
\subsection{Nonlinear, Dynamic Finite Element Model}

The FE model we use was previously developed by~\citet{parkIJSS2012}.  In that work, the corresponding author and collaborators developed a nonlinear, dynamic FEM formulation of the governing nonlinear electromechanical field equations of~\citet{suoJMPS2008} that are summarized in (\ref{eq:suo1}) and (\ref{eq:suo2}).  By using a standard Galerkin FE approximation to both the mechanical displacement and electric potential fields, and incorporating inertial effects in the mechanical momentum equation, an implicit, coupled, monolithic nonlinear dynamic FE formulation was obtained with the governing equations~\citep{parkIJSS2012}
\begin{equation}\label{eq:fe1} \left(\begin{array}{cc}{\Delta\mbf{a}} \\ {\Delta\Phi}\end{array}\right)=-\left(\begin{array}{cc} {\mbf{M}+\beta\Delta t^{2}\mbf{K}^{mm}} & {\mbf{K}^{me}} \\ {\beta\Delta t^{2}\mbf{K}^{em}} & {\mbf{K}^{ee}} \end{array}\right)^{-1}\left(\begin{array}{cc}{\mbf{R}^{mech}} \\ {\mbf{R}^{elec}}\end{array}\right)
\end{equation}
where $\Delta\mbf{a}$ is the increment in mechanical acceleration, $\Delta\Phi$ is the increment in electrostatic potential, $\beta=0.25$ is the standard Newmark time integrator parameter, $\mbf{R}^{mech}$ is the mechanical residual, $\mbf{R}^{elec}$ is the electrical residual, and the various stiffness matrices $\mbf{K}$ include the purely mechanical ($\mbf{K}^{mm}$), mixed electromechanical ($\mbf{K}^{me}=\mbf{K}^{em}$), and purely electrostatic ($\mbf{K}^{ee}$) contributions.  Details regarding the residual vectors and the various mechanical, electromechanical and electrostatic stiffnesses can be found in previous work~\citep{parkIJSS2012}.  

In the present work, volumetric locking due to the incompressible material behavior was alleviated using the Q1P0 method of~\citet{simoCMAME1985}.  While viscoelastic effects have previously been accounted for within the FE model~\citep{parkSM2013}, these effects are neglected in the present work such that the effects of surface tension on electromechanical instabilities in DEs can be investigated without other physical complications.

\subsection{Surface Tension}

The major computational contribution of this work is in adding surface tension effects to the FE formulation of DEs previously developed by~\citet{parkIJSS2012}, such that coupled electro-elasto-capillary effects can be studied.  We discuss the relevant technical details in this section.

\begin{figure} \begin{center}
\includegraphics[scale=0.4]{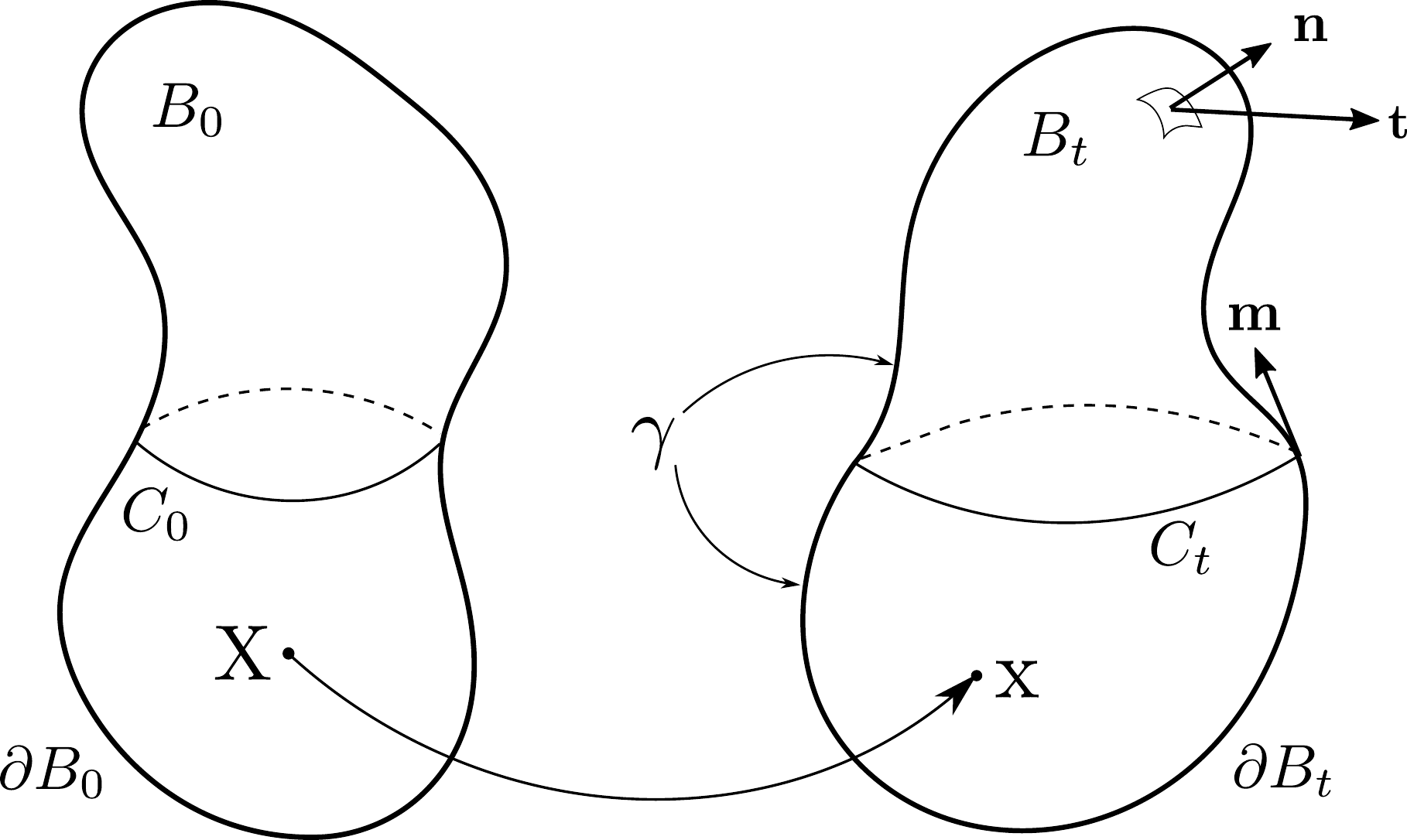}
\caption{The reference body $B_0$ and the deformed body $B_t$ at time $t$.}
\label{ref} \end{center}
\end{figure}

FE formulations of surface tension have been given by~\citet{saksonoCM2006} and~\citet{javiliCMAME2010}, amongst others.  We note that the recently published work of~\citet{henannSM2014} uses the model presented by~\citet{saksonoCM2006}, with an incompressible hyperelastic material model for the deforming solid.  In the present work, we utilize the dynamic formulation of~\citet{saksonoCM2006a}, where the utility and importance of using inertia to capture, using FE, the electromechanical instabilities that occur in DEs was shown in the previous works of~\citet{parkIJSS2012,parkCMAME2013,parkSM2013}.  We now briefly describe the formulation, and the resulting electro-elasto-capillary FE equations that we solve, while a schematic of the current and reference configurations for the large deformation kinematics is shown in Figure (\ref{ref}).

Using the~\citet{saksonoCM2006a} model, we begin with the force continuity at the solid-liquid interface, 
\begin{equation}\label{eq:sak1} {\sigma} \mbf{n}=-p_{ext}\mbf{n}+2H\gamma\mbf{n}
\end{equation}
where $\mbf{n}$ is the unit normal, $\sigma$ is the Cauchy stress, $H$ is the mean curvature, and $p_{ext}$ is an external pressure.  The spatial (updated Lagrangian) form of the weak form of the momentum equation can then be written as
\begin{dmath}\label{eq:sak2} \int (\sigma:\nabla\mbf{w}-\rho(\mbf{b-a})\cdot\mbf{w})dv-\int \mbf{t}\cdot\mbf{w}da-\int (-p_{ext}\mbf{n}\cdot\mbf{w})da \\ \nonumber
+\int (\gamma\nabla_{s}\cdot\mbf{w})da-\int (\gamma\mbf{w}\cdot\mbf{m})ds = 0
\end{dmath}
where $\mbf{w}$ is the virtual displacement, $\mbf{t}$ is the traction, $\mbf{b}$ is the body force, $\nabla_{s}=(\mbf{I}-\mbf{n}\otimes\mbf{n})\nabla$ is the surface gradient operator, $\rho$ is the density and $\mbf{a}$ is the acceleration.  In the present work, we neglect the three-phase contact line, i.e. the integral in (\ref{eq:sak2}) over the line $ds$, for simplicity.

The FE form of (\ref{eq:sak2}) can be obtained by making the usual Galerkin approximation of both the displacements and virtual displacements with the same shape functions, resulting in the (mechanical) residual $\mbf{R}^{mech}(\mbf{X})$
\begin{equation}\label{eq:sak3} \mbf{R}^{mech}(\mbf{X})=\mbf{M}\ddot{\mbf{X}}+\mbf{F}^{int}-\mbf{F}^{ext}+\mbf{F}^{surf}=0
\end{equation}
where the various terms in (\ref{eq:sak3}) take the following form for each element $e$
\begin{eqnarray}\label{eq:sak4} \mbf{M}_{e}=\int\rho\mbf{N}^{T}\mbf{N}dv \\ \nonumber
\mbf{F}_{e}^{int}=\int\mbf{B}^{T}\sigma dv \\ \nonumber
\mbf{F}_{e}^{ext}=\int\mbf{N}^{T}\mbf{b}dv + \int\mbf{N}^{T}\mbf{t}da \\ \nonumber
\mbf{F}_{e}^{surf}=\int\gamma\nabla_{s}\mbf{N}da
\end{eqnarray}
where $\mbf{N}$ and $\mbf{B}$ are the standard FE shape function and gradient, respectively.  As can be seen in (\ref{eq:sak3}), the only non-standard term compared to the standard discretization of the momentum equation arises in the surface force $\mbf{F}^{surf}$, and the corresponding surface stiffness $\mbf{K}^{surf}$.  

\begin{figure} \begin{center}
\includegraphics[scale=0.3]{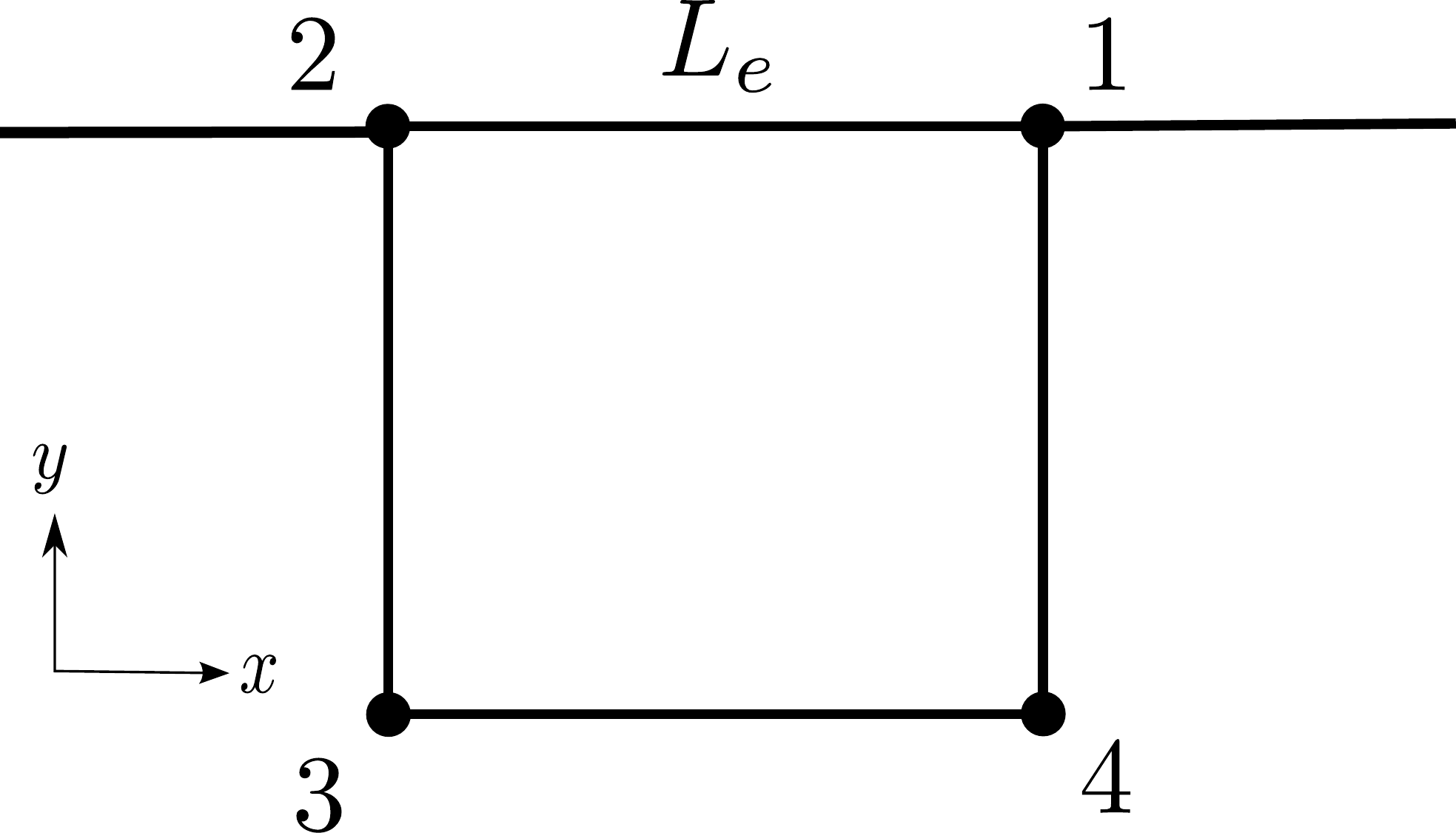}
\caption{Schematic of finite element nodal numbering consistent with surface tension formulation given in Eqs. (\ref{eq:sak5}) and (\ref{eq:sak6}).  Nodes 1 and 2 are surface nodes.}
\label{element} \end{center}
\end{figure}

~\citet{henannSM2014} analytically derived the surface force $\mbf{F}^{surf}$ for a two-dimensional, 4-node bilinear quadrilateral element as
\begin{equation}\label{eq:sak5} \mbf{F}^{surf}=-\frac{h\gamma}{L_{e}}\left(\begin{array}{cccc} {x_{1}-x_{2}} \\ {y_{1}-y_{2}} \\ {x_{2}-x_{1}} \\ {y_{2}-y_{1}} \end{array}\right)
\end{equation} 
The surface stiffness $\mbf{K}^{surf}$ can be obtained through linearization of the surface force $\mbf{F}^{surf}$ in (\ref{eq:sak3}).  This value for a 4-node bilinear quadrilateral element in two-dimensions was also given analytically by~\citet{henannSM2014} as
\begin{dmath} \label{eq:sak6} \mbf{K}^{surf}=\frac{h\gamma}{L_{e}}\left(\begin{array}{cccc} {1} & {0} & {-1} & {0} \\ {0} & {1} & {0} & {-1} \\ {-1} & {0} & {1} & {0} \\ {0} & {-1} & {0} & {1} \end{array}\right)-\frac{h\gamma}{L_{e}^{3}}\left(\begin{array}{cccc} {x_{1}-x_{2}} \\ {y_{1}-y_{2}} \\ {x_{2}-x_{1}} \\ {y_{2}-y_{1}} \end{array}\right)\left(\begin{array}{cccc} {x_{1}-x_{2}} \\ {y_{1}-y_{2}} \\ {x_{2}-x_{1}} \\ {y_{2}-y_{1}} \end{array}\right)^{T}
\end{dmath}
where $L_{e}=\sqrt{(x_2-x_1)^2+(y_2-y_1)^2}$ is the length of the FE face containing nodes 1 and 2.  The ordering of the FE nodes corresponding to Eqs. (\ref{eq:sak5}) and (\ref{eq:sak6}) is shown in Figure (\ref{element}).  

The final coupled electromechanical FE equations we solve, which include the surface tension terms, can be written as
\begin{equation}\label{eq:sak7}\footnotesize{ \left(\begin{array}{cc}{\Delta\mbf{a}} \\ {\Delta\Phi}\end{array}\right)=-\left(\begin{array}{cc} {\mbf{M}+\beta\Delta t^{2}(\mbf{K}^{mm}+\mbf{K}^{surf})} & {\mbf{K}^{me}} \\ {\beta\Delta t^{2}\mbf{K}^{em}} & {\mbf{K}^{ee}} \end{array}\right)^{-1}\left(\begin{array}{cc}{\mbf{R}^{mech}} \\ {\mbf{R}^{elec}}\end{array}\right)}
\end{equation}
In comparing the new FE formulation including surface tension in (\ref{eq:sak7}) to the previous FE equations of~\citet{parkIJSS2012} in (\ref{eq:fe1}), the only changes are the addition of the surface stiffness $\mbf{K}^{surf}$ to the standard mechanical stiffness matrix $\mbf{K}^{mm}$, as well as the surface contribution $\mbf{F}^{surf}$ to the mechanical residual $\mbf{R}^{mech}$, as shown previously in (\ref{eq:sak3}) and (\ref{eq:sak4}).

\section{Numerical Results}

All numerical simulations were performed using the open source simulation code~\citet{tahoe} using standard 4-node, bilinear quadrilateral finite elements within a two-dimensional, plane strain approximation.  Before any application of electrostatic loading via applied voltages, the surface tension is first applied incrementally until the desired value is reached.  This incremental approach is necessary to avoid computational instabilities, as previously discussed by~\citet{javiliCMAME2010}.  The surface tension is applied incrementally by first defining a target value of surface tension $\gamma$, after which we define the following function for the current value of surface tension $\gamma_{0}$: 
\begin{equation} \gamma_{0}=min\left(\gamma,\frac{\gamma t}{t_0}\right)
\label{eq:gammaramp}
\end{equation}
where $t$ is the current time and $t_0$ is the total time allotted to reach the prescribed value for surface tension $\gamma$.  Once the system is in equilibrium with the surface tension, voltage is applied in a monotonically increasing fashion.

\subsection{Single Element Tests}

We first perform a suite of parametric benchmark studies to gain a qualitative understanding of how surface tension impacts the electromechanical behavior of DEs undergoing homogeneous and inhomogeneous deformation.  A single 4-node bilinear quadrilateral element of unit length and height is used for these simulations.  For all numerical simulations, we chose the following constitutive parameters for the Arruda-Boyce model in Eq. (\ref{eq:de2}):  $\mu=\epsilon=1$, $\lambda=1000$ and $N=5.0$, while different values of the surface tension $\gamma$ are chosen. 

\subsubsection{Homogeneous Deformation}

\begin{figure} \begin{center} 
\includegraphics[scale=0.25]{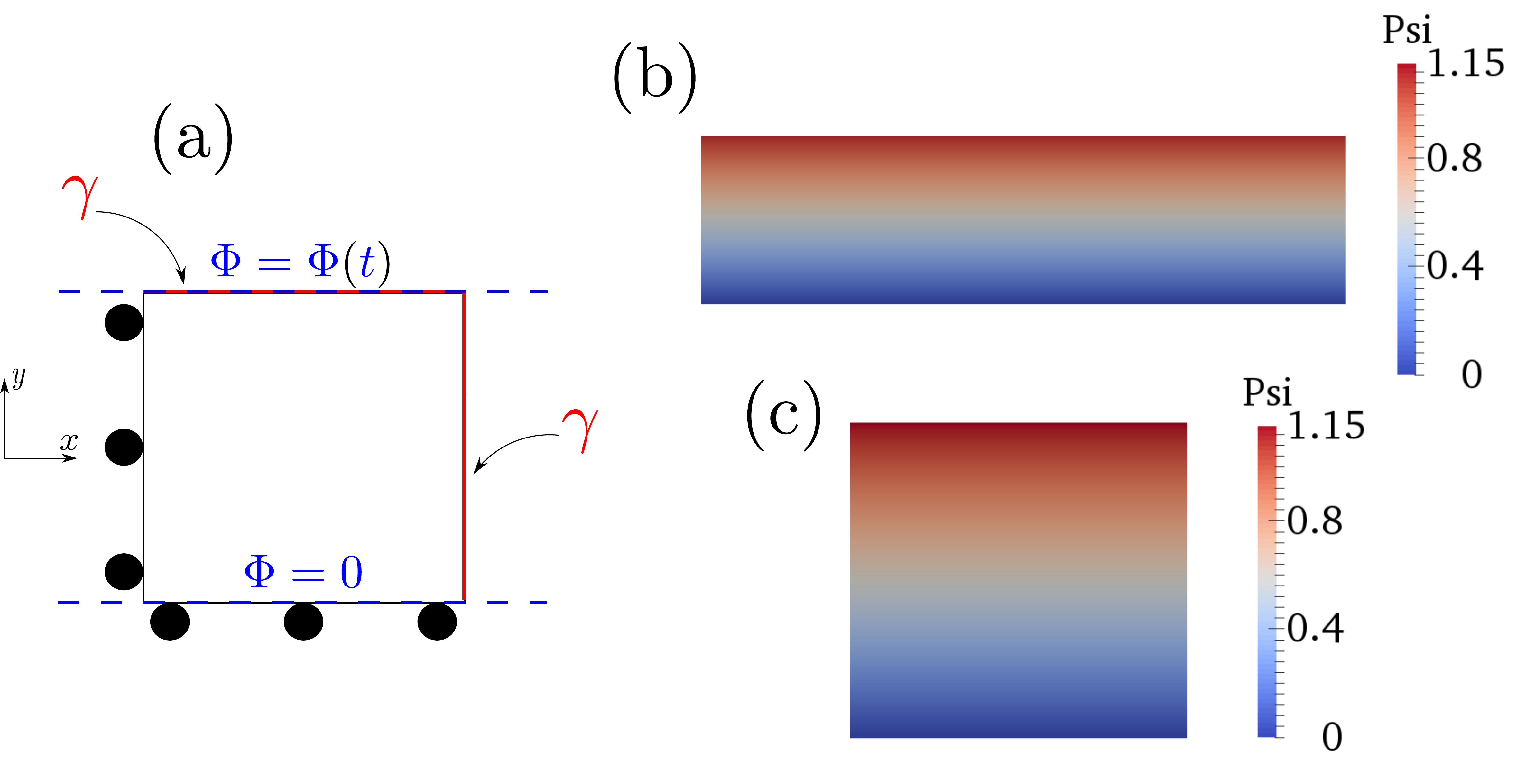}
\caption{(a) Schematic of a single finite element with homogeneous boundary conditions; (b) Deformed configuration with $\gamma/\mu l=0$; (c) Deformed configuration with $\gamma/\mu l=10$.  Note that Psi in (b) and (c) refers to voltage.}
\label{homo3} \end{center} \end{figure}

The first set of boundary conditions are to allow homogeneous deformation of the DE. The electromechanical boundary conditions are as shown in Fig. (\ref{homo3})(a), with rollers on the -x and -y surfaces, and voltage applied at the top surface of the single square element.  Surface tension is present on the free surfaces.  The resulting configurations for elastocapillary number values of $\gamma/\mu l=0$ and $\gamma/\mu l=10$ are shown in Figures (\ref{homo3})(b) and (c).  As can be seen, Figure (\ref{homo3})(b) shows the well-known deformed configuration where the DE contracts along the thickness direction while simultaneously elongating in response to the applied voltage, thus exhibiting the well-known electromechanical snap through instability~\citep{parkIJSS2012}.  However, the configuration in Figure (\ref{homo3})(c) is quite different when $\gamma/\mu l=10$.  Instead of resulting in a rectangular deformed configuration, the single element takes a deformed configuration that is not much changed from the initial, square configuration, which shows that the impact of surface tension is to resist the deformation that would otherwise occur due to the applied voltage.

\begin{figure} \begin{center} 
\includegraphics[scale=1.0]{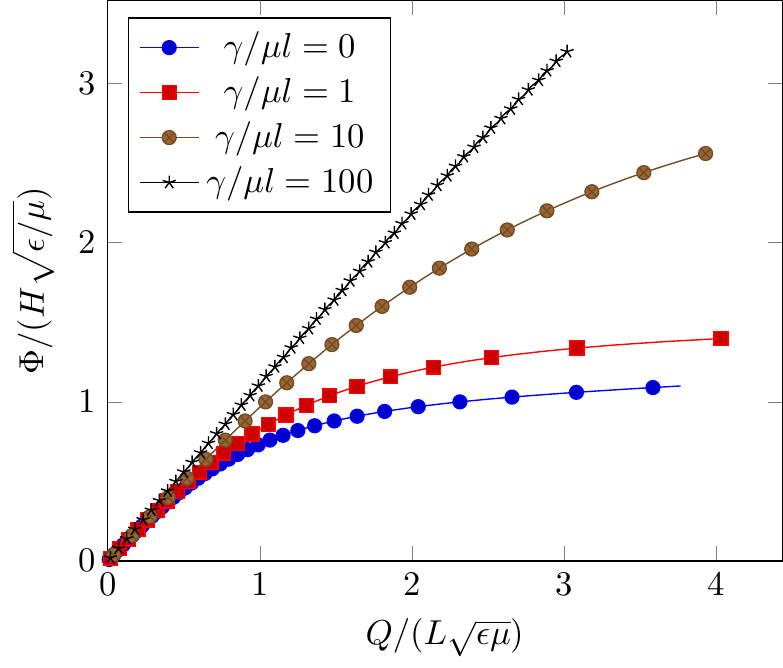}
\caption{Deformation of a single, homogeneously deforming finite element subject to voltage loading.}  
\label{homo} \end{center} \end{figure}

We plot the resulting normalized voltage-charge curves in Figure (\ref{homo}) for various values of $\gamma/\mu l$.  We can see that the effect of increasing surface tension is to significantly increase the voltage that is required to induce the electromechanical instability.  These results are quantitatively in agreement with previous experimental and theoretical studies on soft materials which found that the effect of surface tension is to create a barrier to instability nucleation~\citep{chenPRL2012,moraSM2011}.  We also note that a softening response is not observed in the voltage-charge plot due the fact that a plane strain, and not plane stress~\citep{zhouIJSS2008} approximation in two dimensions is utilized. 

\subsubsection{Inhomogeneous Deformation}

\begin{figure} \begin{center} 
\includegraphics[scale=0.25]{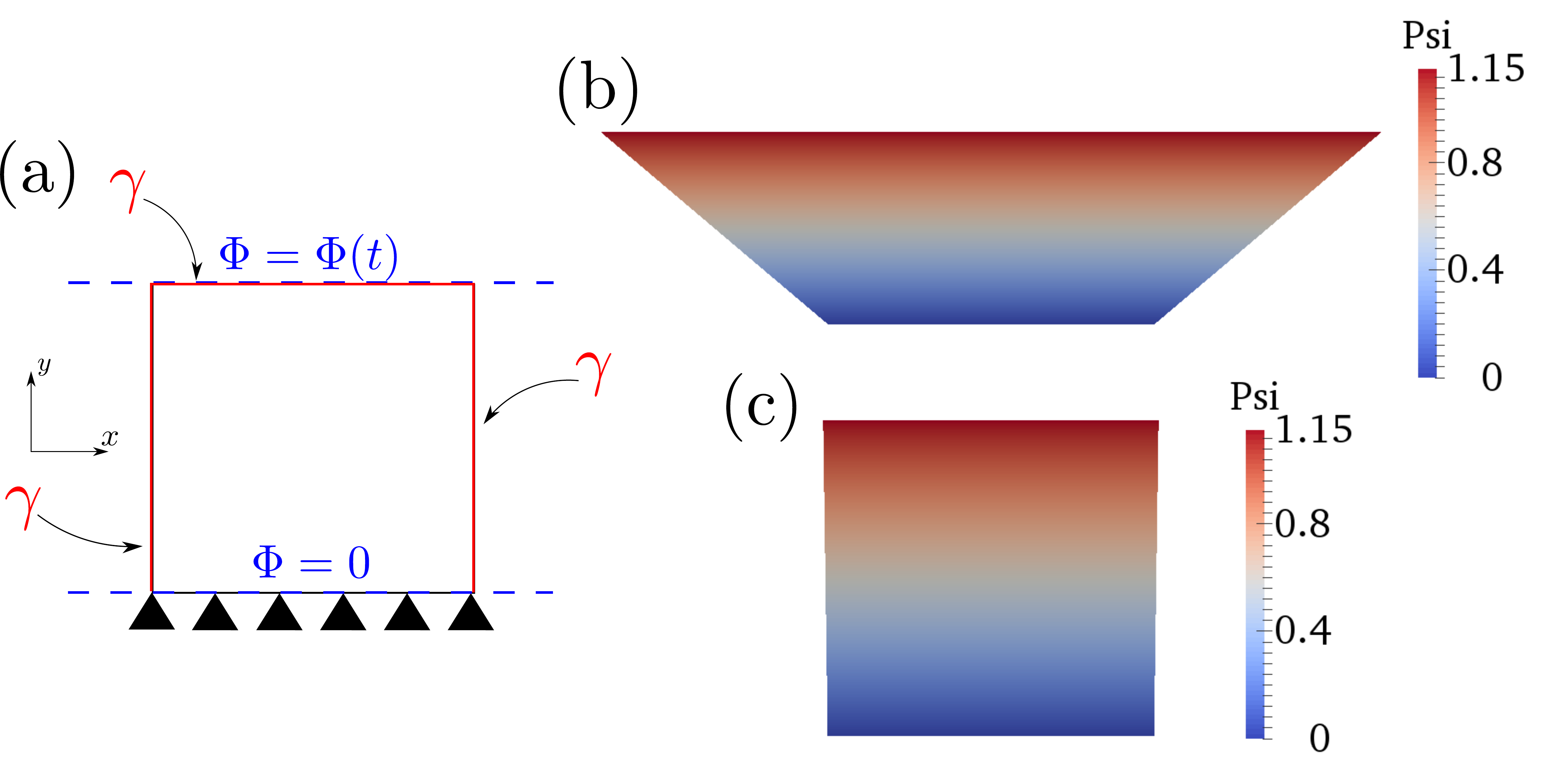}
\caption{(a) Schematic of single finite element with inhomogeneous boundary conditions; (b) Deformed configuration with $\gamma/\mu l=0$; (c) Deformed configuration with $\gamma/\mu l=10$.  Note that Psi in (b) and (c) refers to voltage.}  
\label{inhomo3} \end{center} \end{figure}

Before studying surface tension effects on inhomogeneously deforming DEs, as will be done in subsequent computational examples, we first study a single FE that is fully constrained mechanically along its bottom surface, as illustrated in Figure (\ref{inhomo3})(a), where surface tension is present on the free surfaces.  If no surface tension is present, i.e. $\gamma/\mu l=0$, the top surface of the element bows outward in response to the applied voltage as seen in Figure (\ref{inhomo3})(b).  However, for non-zero surface tension $\gamma/\mu l=10$, the large outward bowing of the top surface that was observed for $\gamma/\mu l=0$ in Figure (\ref{inhomo3})(b) is reduced drastically, as shown in Figure (\ref{inhomo3})(c).  The effect of the surface tension opposing the deformation that is induced by the voltage is similar to that previously observed for the homogeneous boundary conditions.

\begin{figure} \centering 
\includegraphics[scale=1.0]{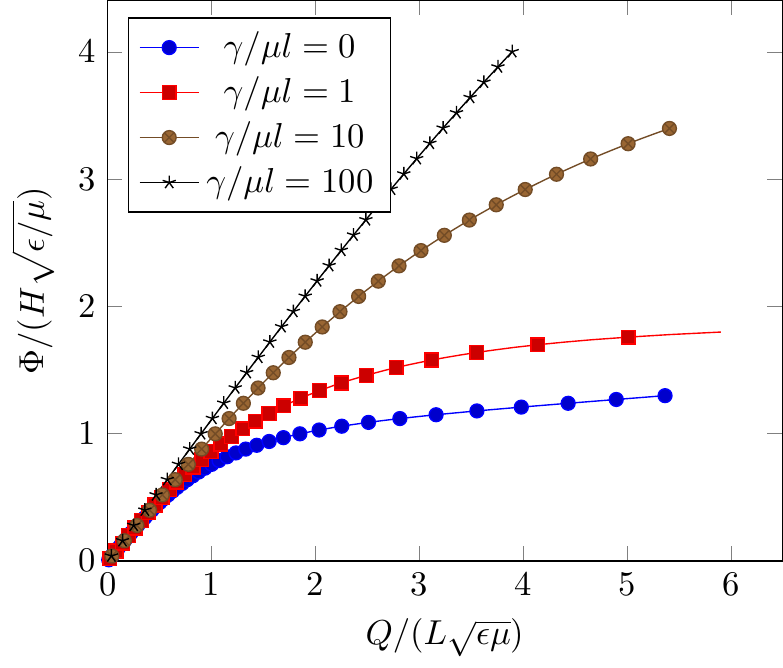}
\caption{Deformation of a single, inhomogeneously deforming finite element subject to voltage loading.}    
\label{inhomo}  \end{figure}


We also analyze the voltage-charge curve for different values of $\gamma/\mu l$ in Figure (\ref{inhomo}).  As can be seen, the major effect of surface tension is again to significantly increase the value of voltage $\Phi$ that is required in order to induce the voltage-induced electromechanical snap-through instability.

\subsection{Surface Creasing in Constrained 2D Strip}

Our next example is based on recent experiments by~\citet{wangPRE2013}, who studied the surface instability mechanisms in constrained DE films in which the top surface is immersed in a fluid.  The fluid was varied such that the top surface of the DE film was subject to different values of surface tension.  Interestingly, upon application of a critical voltage, a transition in the surface instability mechanism from creasing to wrinkling was observed, which was found to be dependent on elastocapillary length $\gamma/(\mu H)$, where $H$ is the film thickness.  Along with the transition in surface instability mechanism, the wavelength of the instability was also found to change, from about $\lambda=1.5H$ for the creasing instability to longer wavelengths, $\lambda=5-12H$ when the elastocapillary length $\gamma/(\mu H)>1$.  

\begin{figure} \begin{center} 
\includegraphics[scale=0.6]{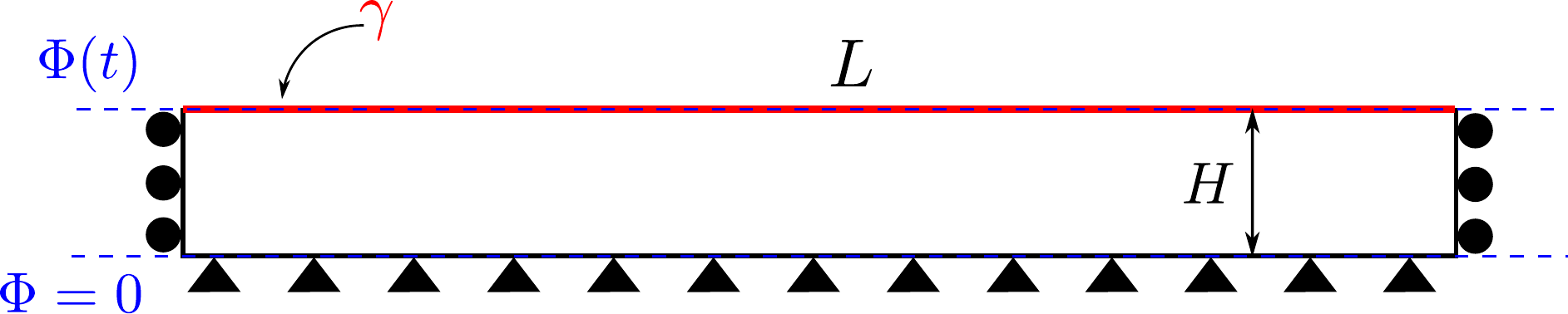}
\caption{Schematic of the DE film with electro-elasto-capillary boundary conditions.}
\label{film} \end{center} \end{figure}

We performed FE simulations of the experiments of~\citet{wangPRE2013}, where the schematic of the problem geometry and the relevant electro-elasto-capillary boundary conditions is shown in Figure (\ref{film}), i.e. the bottom surface of the film is fixed, the left and right sides are on rollers while a voltage is applied to the top surface, where the surface tension is also present.   Most results shown are for a DE film of dimensions 160x4, which was discretized with square 4-node bilinear quadrilateral finite elements having an edge length of unity.  

\begin{figure} \begin{center} 
\includegraphics[scale=0.35]{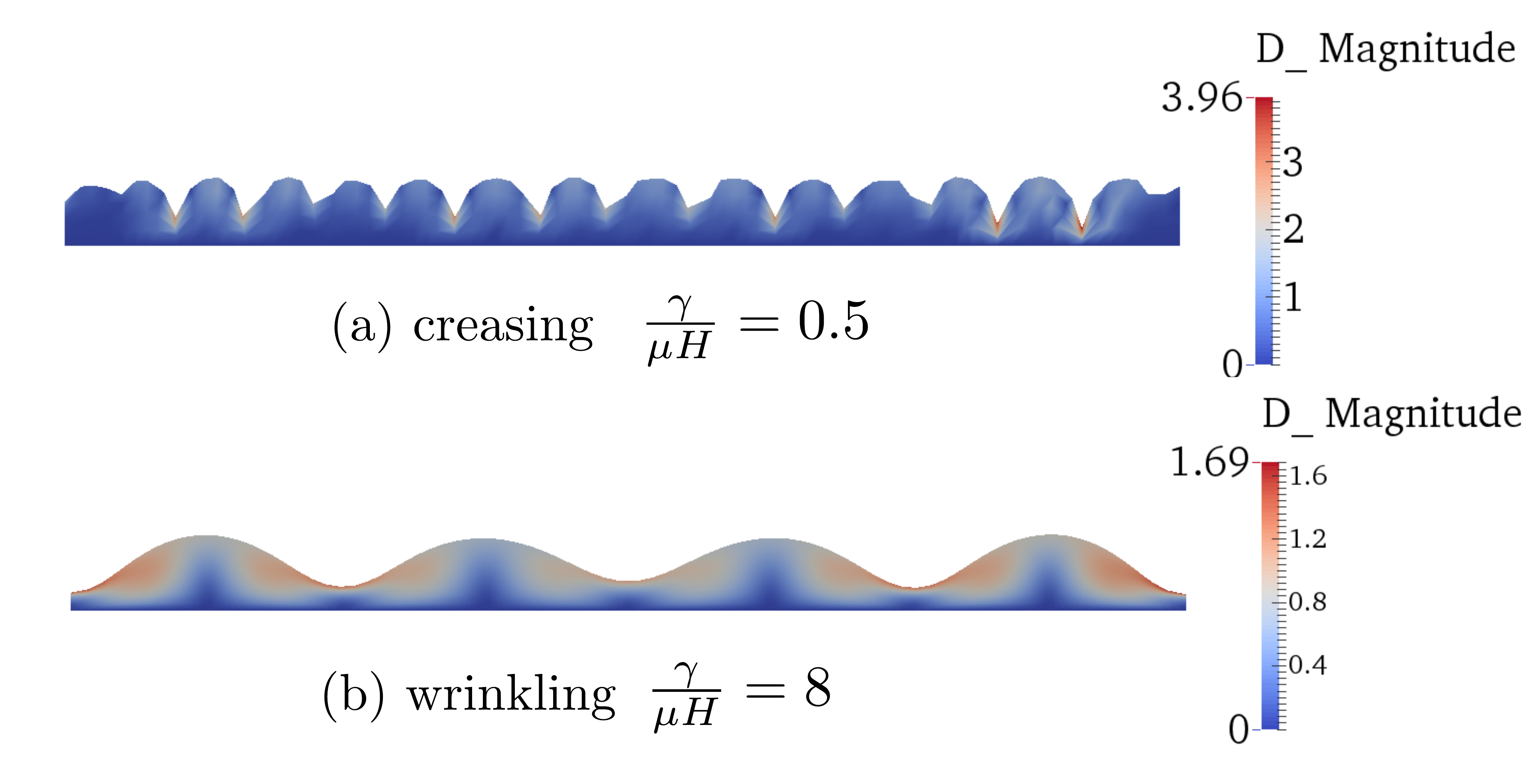}
\caption{Computationally observed transition in the surface instability mechanism in DEs as a function of the elastocapillary length $\gamma/(\mu H)$, for a DE film of dimensions 80x4.  The variable D_Magnitude refers to the magnitude of the displacement field.}
\label{wrinkcreas} \end{center} \end{figure}

We show in Figure (\ref{wrinkcreas})(a) the difference in surface instability mechanism depending on the elastocapillary length.  As can be seen, when the elastocapillary length is small (i.e. $\gamma/\mu H=0.5$), the surface instability mechanism is that of creasing, or localized folds, as previously observed both experimentally~\citep{wangPRL2011a}, and computationally~\citep{parkCMAME2013}.  As the elastocapillary length increases to become similar to the film height, the surface instability mechanism changes to wrinkling, as shown in Figure (\ref{wrinkcreas})(b).  The change in instability mechanism is characterized by a significantly larger instability wavelength as compared to the creasing instability in Figure (\ref{wrinkcreas})(a).  Furthermore, rather than abrupt, localized folds as in the creasing instability in Figure (\ref{wrinkcreas})(a), the surface exhibits a more gradual and undulating pattern as seen in Figure (\ref{wrinkcreas})(b).  

\begin{figure} \centering 
\includegraphics[scale=1.0]{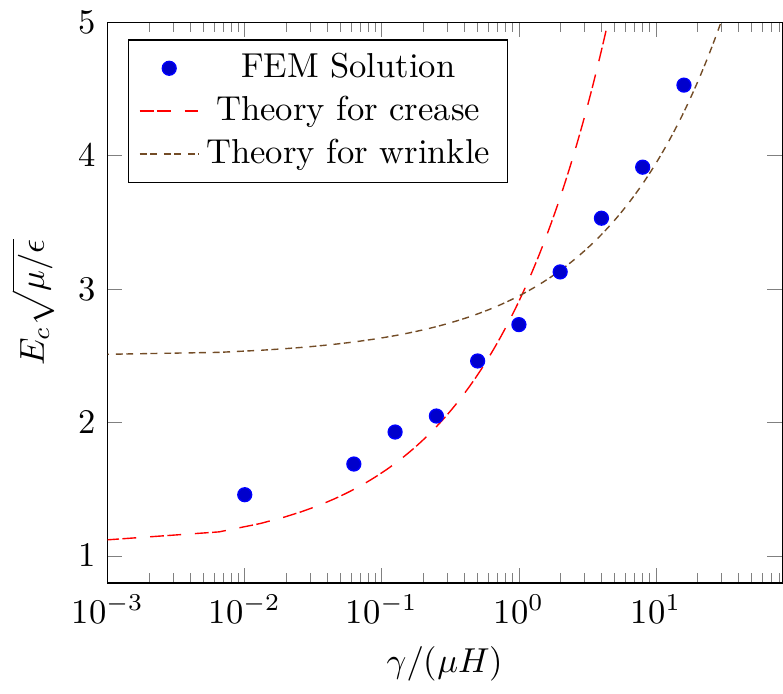}
\caption{Normalized critical electrical field as a function of elastocapillary length for a 160x4 DE film.}   
\label{ecrit2} \end{figure}


This instability transition was also characterized experimentally by plotting the normalized electric field $E_{c}/\sqrt{\mu/\epsilon}$  as a function of the elastocapillary length, as shown in Figure (\ref{ecrit2}), where $E_{c}$ is the value of the electric field when the instability nucleates.  The analytic solutions for the critical electric fields to nucleate wrinkles and creases were developed by~\citet{wangPRE2013}, and are written as:  
\begin{equation} \label{eqn:ECcrease}
E^{c}_{crease}\approx1.03\sqrt{\frac{\mu}{\epsilon}}+1.88\sqrt{\frac{\gamma}{H\epsilon}}
\end{equation}
\begin{equation} \label{eqn:ECwrinkle}
E^{c}_{wrinkle}\approx2.49\sqrt{\frac{\mu}{\epsilon}}+0.46\sqrt{\frac{\gamma}{H\epsilon}}
\end{equation}
As seen in Figure (\ref{ecrit2}), the FE model is able to capture the general trends, including the transition in the value of the critical electric field $E_{c}$ predicted theoretically.  The FE prediction for the critical electric field is generally larger than the analytic theory, which is expected given that while a locking-resistant Q1P0 formulation was used~\cite{simoCMAME1985}, the FE-discretized structure is still stiffer than the continuum.  

\begin{figure} \centering 
\includegraphics[scale=1.0]{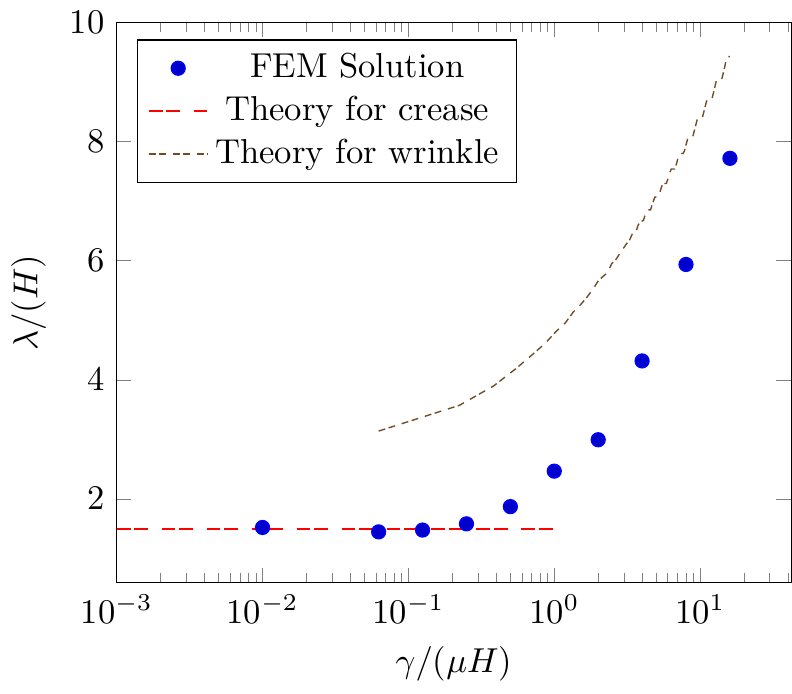}
\caption{Normalized instability wavelength as a function of elastocapillary length for a 160x4 DE film.}
\label{ecrit1} \end{figure}


We also calculated the normalized instability wavelength $\lambda/H$ as a function of the elastocapillary length as shown in Figure (\ref{ecrit1}).  In that figure, the wrinkling wavelength as a function of elastocapillary number is given by
\begin{equation} \label{eqn:wrinkle}
\frac{\epsilon E^2}{\mu}=2Hk\frac{1+2e^{2Hk}+e^{4Hk}+4e^{2Hk}H^2k^2}{-1+e^{4hk}-4e^{2Hk}Hk}+\left(Hk\right)^2\frac{\gamma}{\mu H}
\end{equation}
The creasing wavelength is fixed as $\lambda_{crease}=1.5H$, which was previously obtained by~\citet{wangPRE2013}.

For elastocapillary numbers $\gamma/(\mu H)<1$, we find the normalized wavelength $\lambda/H$ to be close to the value of 1.5 predicted theoretically~\cite{wangPRE2013}.  For larger elastocapillary numbers, i.e. $\gamma/(\mu H)>1$, where wrinkling is observed, a dramatic increase in normalized instability wavelength $\lambda/H$ is observed.  While our FE simulations capture the instability wavelength transition, the predicted wavelength after the transition to wrinkling occurs is smaller than the analytic theory.  

This discrepancy was also observed in the experimental studies of~\citet{wangPRE2013}.  In that work, like the current FE models, the critical electric fields for creasing and wrinkling as in Figure (\ref{ecrit2}) were in better agreement with the analytic theory than the instability wavelength.  Specifically, the experiments also found a smaller instability wavelength in the transition region from creasing to wrinkling than the analytic solution.  The FE predictions do fall within the bounds observed experimentally by~\citet{wangPRE2013}, though at the lower end of the observed wavelengths.  This is likely due to differences in how the instability wavelength was calculated in our plane strain 2D FE simulations as compared to the 3D experimental studies.  In particular, there is larger spatial variation in the wavelengths observed in the 3D experimental structure, whereas the 2D plane strain approximation results in a constant instability wavelength through the thickness of the film, and so we expect and find that the FE predictions fall towards the lower end of the experimental values.

\subsection{Bursting drop in a solid dielectric}

\begin{figure} \begin{center} 
\includegraphics[scale=0.35]{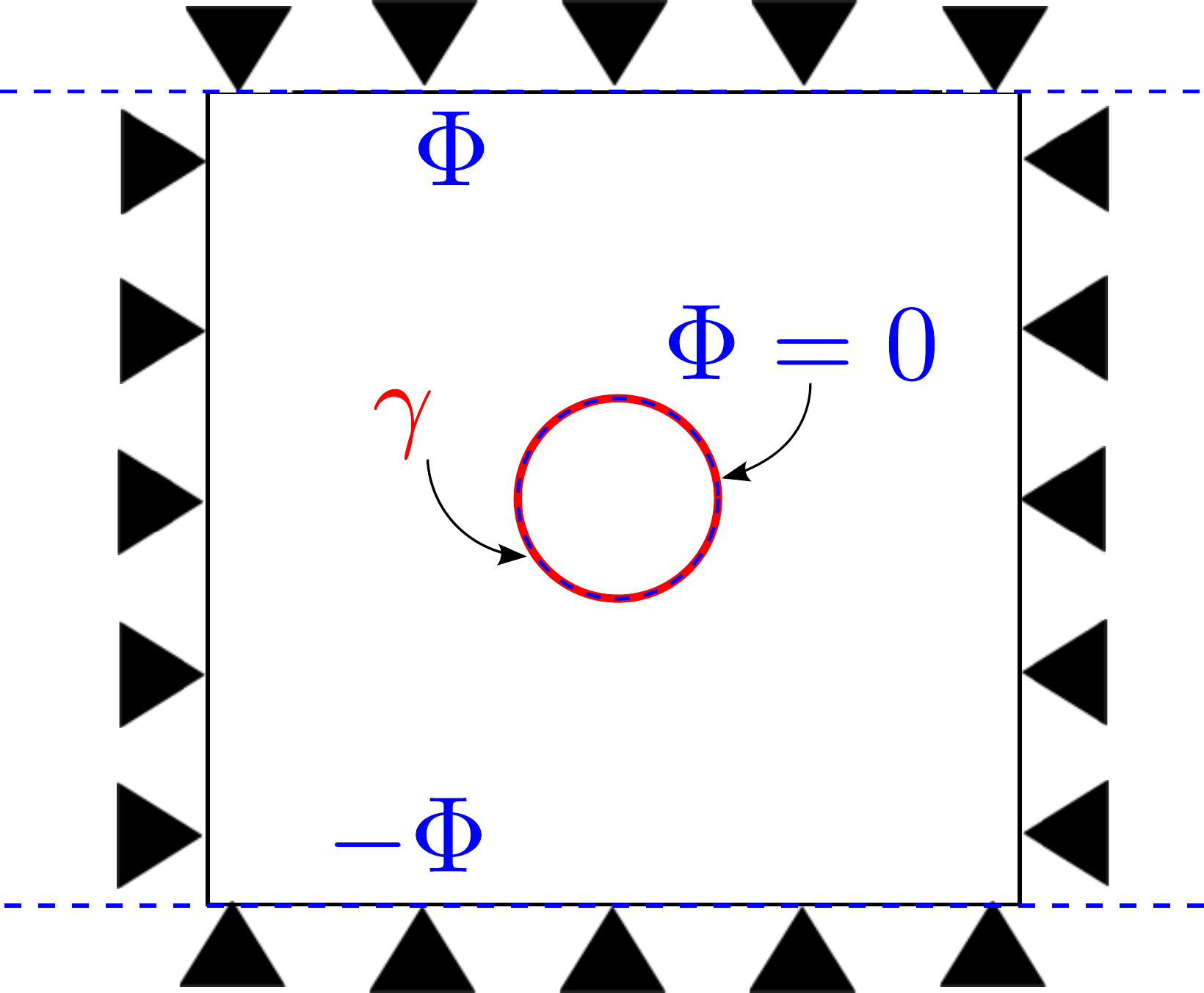}
\caption{Computational model for bursting drop in dielectric elastomer. }  
\label{bursting2} \end{center} \end{figure}

Our final numerical example is a computational study of the experiments of~\citet{wangNC2012}.  The experimental configuration is shown in Figure (\ref{bursting2}).  The problem is one of a dielectric film with a small hole containing a conductive liquid, for example NaCl solution.  All edges of the film are constrained mechanically, and a voltage differential $\Phi$ is applied across the film.  The novel experimental finding was the first observation of instabilities of drops in solids, in which the drop in the center of the film begins to elongate in a crack-like fashion towards the boundaries at which the voltage is applied.  While viscoelastic effects on the bursting drop instability were previously studied by~\citet{parkSM2013}, what has not been investigated, either experimentally or theoretically, is the effect of changing the conductive fluid within the hole.  We account for the effect of different conductive fluids by changing the surface tension around the hole perimeter, as shown schematically in Figure (\ref{bursting2}).

The 2D plane strain model in Figure (\ref{bursting2}) had dimensions of 20 by 20, with the radius of the hole being 2.  The geometry was discretized using 4-node bilinear quadrilateral finite elements with a mesh spacing of 0.25, for a total number of 6271 elements and 6456 nodes.

\begin{figure}
\centering
\includegraphics[scale=0.48]{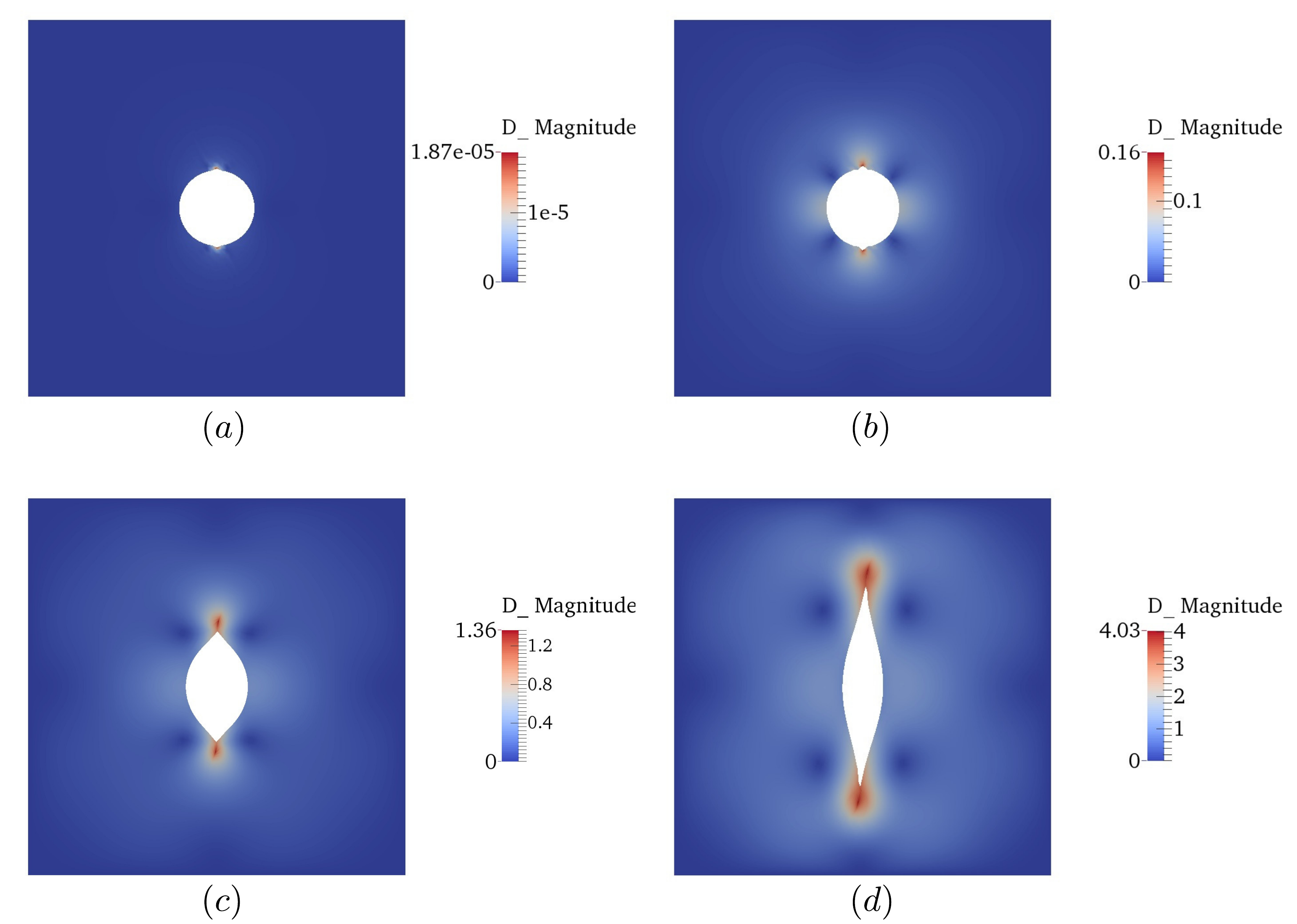}
\caption{Representative snapshots of \hsp{bursting drop evolution} in constrained DE film. }
\label{bursting1} \end{figure}

\begin{figure}
\centering
\includegraphics[scale=0.3]{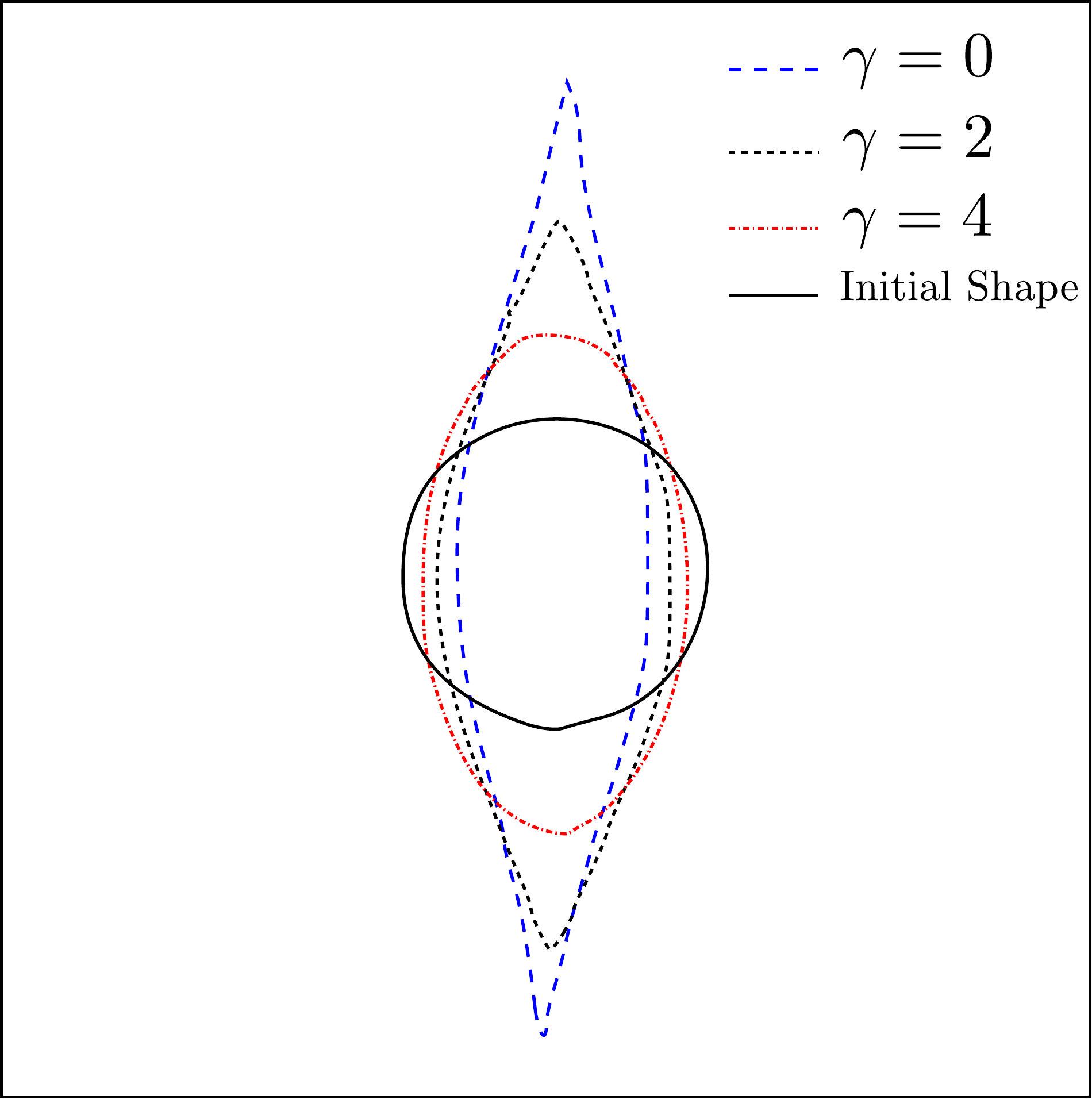}
\caption{Snapshot of bursting drop geometry taken at same time for different values of surface tension $\gamma$.}
\label{bursting3} \end{figure}

The electromechanical instability is shown in Figure (\ref{bursting1}).  There, it is seen that as the applied voltage increases, the circular drop begins to change shape, and then as shown in Figures (\ref{bursting1})(b)-(d), begins to elongate in the direction of the applied voltage, where the elongation resembles crack-like propagation.  

We quantify the effects of surface tension on the nature of the bursting drop.  We show in Figure (\ref{bursting3}) three snapshots, overlaid on top of each other, of the bursting drop geometry taken at the same time for different values of surface tension $\gamma$.  As can be seen, the drop has elongated the most for the case without surface tension, i.e. $\gamma=0$ as compared to its initial, circular shape.  As the surface tension increases, the bursting drop elongation becomes progressively more delayed, which again shows that surface tension acts as a barrier to instability nucleation in electroactive polymers.  \hsp{We also note that the slightly asymmetric direction of bursting drop elongation is due to the fact that bursting drop follows the contours of the FE mesh, which is slightly asymmetric with respect to the hole.}

\begin{figure}
\centering
\includegraphics[scale=1.0]{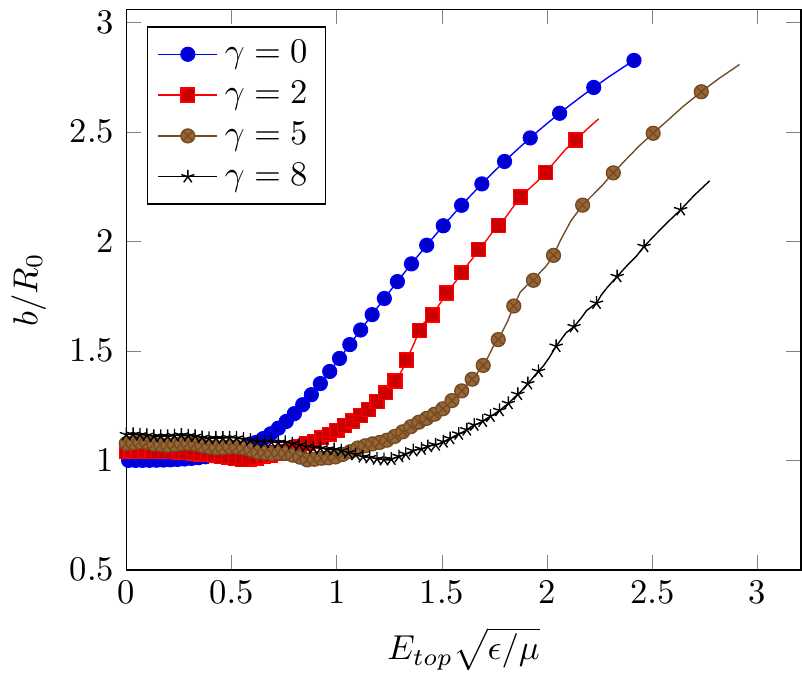}
\caption{Position of bursting drop tip as a function of applied electric field $E_{top}$, where $R_{0}$ is the initial radius of the drop, and $b$ is the long axis of the bursting drop.}
\label{crack} \end{figure}

\hsp{It is also of interest to quantify the position of the bursting drop tip as a function of applied electric field, which is shown in Figure (\ref{crack}).  As can be seen, the nucleation for larger values of surface tension is delayed, and thus occurs at larger values of applied electric field $E_{top}$.  Once the nucleation occurs, and the bursting drop begins to elongate towards the applied electric field, it appears as though the surface tension does not have a significant impact on the elongation rate of the bursting drop tip, which is in contrast to the effect that viscoelasticity was previously observed to have~\citep{parkSM2013}.  This is likely because viscoelasticity impacts the stiffness of the entire DE geometry, whereas the surface tension creates a nucleation barrier only at the bursting drop tip.}

\section{Conclusions}

We have presented a new dynamic, finite deformation finite element model of dielectric elastomers that incorporates surface tension to capture elastocapillary effects on the electromechanical deformation.  The simulations demonstrate that increasing surface tension, or equivalently the elastocapillary number, results in an increase in the critical voltage or electric field needed to nucleate an electromechanical instability in the dielectric elastomer.  We also demonstrated a transition in surface instability mechanism from creasing to wrinkling in constrained dielectric elastomer films by increasing the elastocapillary number.  The present results indicate that the proposed methodology may be beneficial in studying the electromechanical deformation and instabilities for dielectric elastomers in the presence of surface tension.

\section{Acknowledgements}

HSP and SS acknowledge funding from the ARO, grant W911NF-14-1-0022.

\section{Appendix}

\begin{figure} \begin{center} 
\includegraphics[scale=1.0]{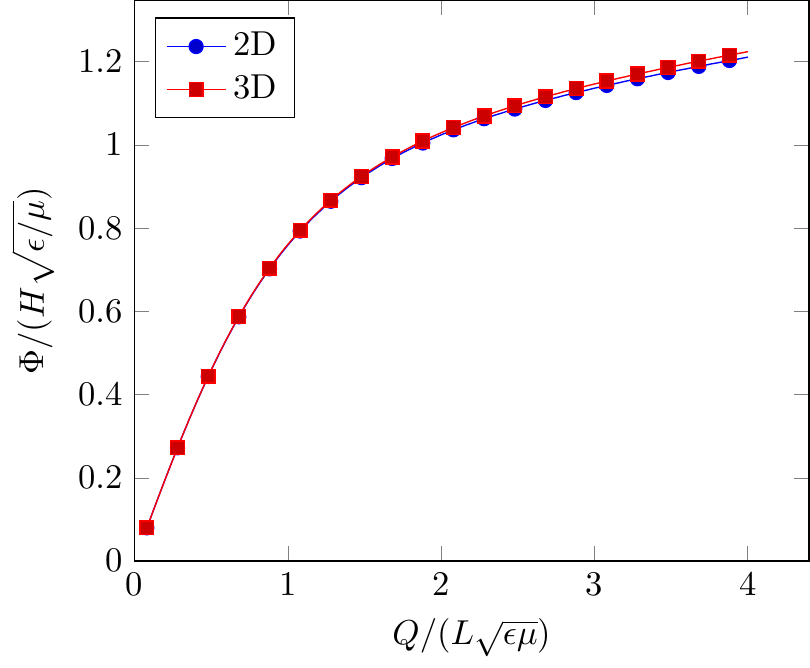}
\caption{Comparison of 2D and constrained 3D results for inhomogeneous deformation of a single finite element.}
\label{2d3d} \end{center} \end{figure}


We show here verification of the 2D Q1P0 formulation in alleviating volumetric locking.  To do so, we compare it to results obtained using a single 8-node hexahedral element with the previously published Q1P0 formulation of~\citet{parkCMAME2013}.  The 2D results were obtained using the 2D version of the Q1P0 formulation of~\citet{simoCMAME1985}; the boundary conditions on the single 4-node quadrilateral 2D finite element were the same as those previously seen in Figure (\ref{inhomo3})(a).  For the 3D element, in addition to having its bottom surface completely constrained, all $z$-displacements were set to zero to mimic a 2D plane strain problem.  

The results are shown in Figure (\ref{2d3d}).  As can be seen, the 2D and constrained 3D formulations give quite similar results, validating the present 2D formulation.

\bibliographystyle{model2-names}
\bibliography{biball}

\end{document}